\documentclass[10pt,conference]{IEEEtran}
\IEEEoverridecommandlockouts
\usepackage{cite}
\usepackage{amsmath,amssymb,amsfonts}
\usepackage{algorithmic}
\usepackage{graphicx}
\usepackage{textcomp}
\usepackage{xcolor}
\usepackage{url}
\usepackage[english]{babel}
\usepackage{blindtext}

\def\BibTeX{{\rm B\kern-.05em{\sc i\kern-.025em b}\kern-.08em
    T\kern-.1667em\lower.7ex\hbox{E}\kern-.125emX}}

\makeatletter
\newcommand{\linebreakand}{%
  \end{@IEEEauthorhalign}
  \hfill\mbox{}\par
  \mbox{}\hfill\begin{@IEEEauthorhalign}
}
\makeatother

\begin{document}

\title{Towards Decoding Developer Cognition in the Age of AI Assistants}

\author{\IEEEauthorblockN{Ebtesam Al Haque}
\IEEEauthorblockA{\textit{Department of Computer Science} \\
\textit{George Mason University}\\
Fairfax, VA \\
ehaque4@gmu.edu}
\and
\IEEEauthorblockN{Chris Brown}
\IEEEauthorblockA{\textit{Department of Computer Science} \\
\textit{Virginia Tech}\\
Blacksburg, VA \\
dcbrown@vt.edu}
\and
\IEEEauthorblockN{ Thomas D. LaToza}
\IEEEauthorblockA{\textit{Department of Computer Science} \\
\textit{George Mason University}\\
Fairfax, VA \\
tlatoza@gmu.edu}
\linebreakand 
\IEEEauthorblockN{Brittany Johnson}
\IEEEauthorblockA{\textit{Department of Computer Science} \\
\textit{George Mason University}\\
Fairfax, VA \\
johnsonb@gmu.edu}
}

\maketitle

\begin{abstract}
 
% III. Structured Abstract (required) The abstract should describe the following in 200 words or so:

% -Background/Context: What is your research about? Why are you doing this research, why is it interesting?
\textbf{Background:} The increasing adoption of AI assistants in programming has led to numerous studies exploring their benefits. While developers consistently report significant productivity gains from these tools, empirical measurements often show more modest improvements. While prior research has documented self-reported experiences with AI-assisted programming tools, little to no work has been done to understand their usage patterns and actual cognitive load imposed in practice.

% indicating that there may be other factors influencing developers’ perceptions of productivity.
% Example: “The enterprise is the flag ship of the federation, and it allows families to travel onboard. However, there are no studies that evaluate how this affects the crew members.”

% -Objective/Aim: What exactly are you studying/investigating/evaluating? What are the objects of the study? We welcome both confirmatory and exploratory types of studies.
\textbf{Objective:} In this exploratory study, we aim to investigate the role AI assistants play in developer productivity. Specifically, we are interested in how developers' expertise levels influence their AI usage patterns, and how it impacts their actual cognitive load and productivity during development tasks. We also seek to better understand how this relates to their perceived productivity.  
% Example (Confirmatory): We evaluate whether the frequency of sick days, the work effectiveness and efficiency differ between science officers who bring their family with them, compared to science officers who are serving without their family.

% Example (Exploratory): We investigate the problem of frequent Holodeck use on interpersonal relationships with an ethnographic study using participant observation, in order to derive specific hypotheses about Holodeck usage.

% -Method: How are you addressing your objective? What data sources are you using?
\textbf{Method:} We propose a controlled observational study combining physiological measurements (EEG and eye tracking) with interaction data to examine developers' use of AI-assisted programming tools. We will recruit professional developers to complete programming tasks with and without AI assistance while we measure their cognitive load and task completion time. Through pre- and post-task questionnaires, we will collect data on perceived productivity and cognitive load using NASA-TLX. 
% Example: We conduct an observational study and use a between subject design. To analyze the data, we use a t-test or Wilcoxon test, depending on the underlying distribution. Our data comes from computer monitoring of Enterprise crew members.

\end{abstract}


\section{Introduction}
% [AI4SE intro sentence]. We find that developers measure usually their productivity in terms of task completion time... Prior studies explore how developers use AI and its impact on their productivity. Perceived productivity benefits are higher than empirically measured productivity. Studies that report on cognitive load only report perceived cognitive load. 

% This study aims to investigate the relationship between perceived and actual productivity gains when using AI assistants for information seeking in software development tasks, compared to traditional sources. We will examine how developers' prior knowledge influences their success with AI assistants and explore the cognitive load implications across... In addition, we will analyze the validation efforts required...
In recent years, we have seen a significant increase in the development and use of tools powered by artificial intelligence (AI). 
This includes, but is not limited to, AI-assisted tools for code generation~\cite{chen2021evaluating}, information seeking ~\cite{haque2024information}, and planning ~\cite{bairi2024codeplan}.
In software engineering, most of the research and development efforts have focused on AI-assisted tools that can support developers' efficiency in writing high quality code.
Given the increased usage of AI for code generation, many research efforts have attempted to provide insights into the ability for these tools to support developers in practice~\cite{parnin2023building, 10.1145/3597503.3608128, haque2024information, chen2021evaluating, fan2023large}.

% \bj{May need more here to fill in the narrative around why this work is relevant and important.}
While some research shows significant improvements in task completion times, others indicate that the increased code output comes with tradeoffs~\cite{mozannar2024realhumaneval}. The time saved in writing code is often balanced by new cognitive demands as developers must carefully review and validate AI-generated suggestions~\cite{mozannar2024reading}. Traditional productivity metrics, such as lines of code or task completion speed, fail to capture the shifts in how developers work when collaborating with AI assistants. 

While there exists numerous studies on the technical capabilities, quality, and perceived productivity benefits of AI-assisted code generation, few provide insights into the ways in which perceived benefits, cognitive load, and user expertise impact actual productivity and task completion when using these tools. Furthermore, most existing studies focus on either analysis of existing developer tool use data or experiments that evaluate generated code outside of its use context. 
As a result, there remains a gap in understanding the impact of AI assistants on developers' cognitive processes and decision-making patterns, and how these changes ultimately impact software quality, development practices and developer productivity.
% \bj{$<$-- This sentence may need some work.} 
To this end, we propose a controlled observational study designed to collect multi-dimensional data on developer use of AI-assisted coding tools for completing coding tasks.

% While numerous studies focus on the technical capabilities of AI coding assistants or their immediate task completion metrics and perceived productivity benefits, very few examine actual productivity benefits and its relationship to perceived benefits, cognitive load, and user expertise [needs rephrasing].

\section{Background and Related Work}
% Papers to discuss:
% AI+SE:
% * reading between the lines
% * real human eval
% * github copilot study
% * ..?
% * 
% Cognitive load measurement/other Psycho-Physiological Measures for SE:
% * Can EEG Be Adopted as a Neuroscience Reference for Assessing Software Programmers’ Cognitive Load?
% * Measuring the Cognitive Load of Software Developers: A Systematic Mapping Study

Studies measuring the productivity impact of AI coding assistants have shown mixed but generally positive results. In a controlled experiment with 95 professional developers, Peng et. al.~\cite{peng2023impact} found that access to GitHub Copilot led to a 55.8\% reduction in task completion time for implementing an HTTP server, with less experienced and older developers benefiting the most. Vaithilingam et al. ~\cite{vaithilingam2022expectation} on the other hand, did not find any statistically significant difference in the time needed for task completion. Mozannar et al.~\cite{mozannar2024reading} found that while developers perceive AI tools to be beneficial, they introduce new bottlenecks for reading, understanding and validating AI-generated code suggestion. Adding to this, Imai~\cite{imai2022github} found that while Copilot increased code output compared to human pair programming, the generated code actually required more subsequent modification. Another study found that developers' perceived helpfulness of AI tools did not always align with actual productivity benefits~\cite{vaithilingam2022expectation}.

Prior work has explored various approaches to measuring cognitive load in software engineering contexts. A systematic mapping study by Gonçales et al.~\cite{gonccales2019measuring} identified 33 studies investigating cognitive load measurement in software engineering, finding that 55\% of studies utilized electroencephalography (EEG), while 36\% combined multiple sensors like eye tracking and electrodermal activity for improved accuracy. Fritz et. al.~\cite{fritz2014using} used a combination of EEG, eye tracking and electrodermal activity sensors to predict task difficulty with upto 84\% precision. Siegmund ~\cite{10.1145/3106237.3106268} investigated cognitive load during code comprehension tasks using fMRI and found distinct activation patterns in regions related to working memory, attention, and language processing. 

Despite growing evidence that AI coding assistants introduce new cognitive demands around reviewing and validating suggestions, there has been limited investigation of the cognitive load imposed by programming with AI assistance. Traditional productivity metrics like time-to-completion or lines of code written may overlook the mental effort required to effectively engage with AI tools. The few studies examining cognitive aspects have relied primarily on self-reported measures, which can be unreliable for assessing cognitive load.

% \eah{From reviews: [DONE] One closely related work is Tang et al.'s "A Study on Developer Behaviors for Validating and Repairing LLM-Generated Code Using Eye Tracking and IDE Actions" (arXiv:2405.16081, accepted at VL/HCC 2024). It would be valuable for the authors to discuss how their work differs from and complements the VL/HCC study, as the two appear to have related yet distinct focuses.}

This presents an important gap in understanding how AI assistants affect developers' cognitive load during programming tasks. Psycho-physiological measures like EEG, which have been validated for measuring cognitive load in traditional programming contexts, could provide valuable insights into the mental demands of working with AI assistants. Recently, Tang et al.~\cite{tang2024developer} conducted a focused study on how developers validate and repair LLM-generated code, finding that awareness of AI authorship affected both behavior and cognitive load. While their work provided valuable insights into validation strategies, our study takes a broader view - examining how AI assistance impacts overall developer productivity across different expertise levels, and investigating the relationship between perceived productivity and cognitive load.
This understanding is crucial for designing AI tools that better support developer productivity and wellbeing.

\section{Research Question and Hypotheses}
The goal of our proposed study is to better understand how developers use AI-assisted coding tools in practice and the impact various factors have on the ability to effectively and confidently use these tools to complete coding tasks. 
Broadly, this research is interested in the following research question: 
\textit{How do human factors impact actual productivity and outcomes when using AI assistants for coding?}
For the proposed study, we focus on \textbf{perceived productivity gains}, \textbf{cognitive load}, and \textbf{expertise} as human factors that may be related to or impact productivity and outcomes in practice given insights provided from prior work regarding perceptions of increased productivity, reduced cognitive load, and a need for relevant expertise when using AI-assisted tools to complete software development tasks\cite{haque2024information}\cite{peng2023impact}.

% Rationale H1: Based on the effort justification paradigm, users tend to overvalue solutions they invested more effort in obtaining. Higher cognitive load during AI interaction may lead to inflated perceptions of productivity gains.
    % *talk about productivity paradox

% \textit{Prior work has shown that less experienced developers benefit the most from tools like GitHub Copilot~\cite{peng2023impact}}
    % Rationale H3: AI assistants can provide contextual, natural language explanations and examples, reducing the cognitive overhead of parsing technical documentation in unfamiliar domains.

    % Rationale H4: The nondeterministic nature of AI responses and their potential to hallucinate creates a need for additional validation steps compared to traditional documentation

From our overarching research question, there are four hypotheses we aim to evaluate:
% \eah{From reviews: - All hypotheses need further justification for their design.
% - A more extensive discussion on confounders needs to be discussed.}
% \eah{reviews: - Somewhat minor, but the hypotheses should be split so that each of them is actually a single hypotheses. Right now, H1 and H2 are both describing two hypotheses (through an AND conjunction). Similarly, they should probably be formulated as null hypotheses. Ultimately, this is primarily a cosmetic issue for me.}
% \begin{description}
%     \item[\textbf{H1:}]Developers using AI assistants will report higher perceived productivity gains compared to developers using traditional methods for similar information seeking tasks.
%     \textit{Rationale: The immediate, contextualized responses from AI assistants may create a stronger perception of productivity improvement compared to traditional documentation navigation, but fail to account for the new bottlenecks such as prompt crafting and validating AI responses.}
%     \item[\textbf{H2:}]Developers using AI assistants will report lower cognitive load compared to developers using traditional methods for similar information seeking tasks.
%     \textit{Rationale: The immediate, contextualized responses from AI assistants may create a stronger perception of productivity improvement compared to traditional documentation navigation, but fail to account for the new bottlenecks such as prompt crafting and validating AI responses.}

% \end{description}
\begin{description}
\item[\textbf{H1: Perceived vs. Actual Productivity and Cognitive Load}]
    \item[\textbf{H1a.}] Developers perceive higher productivity gains than their actual measured productivity improvements when using AI assistants for information seeking tasks.\\
    \textit{Rationale: The effort justification paradigm~\cite{festinger1957theory} suggests that higher investment in obtaining solutions may lead to inflated perceptions of their value. This hypothesis may help us understand potential disconnects between perceived and actual benefits of AI assistance.}
    \item[\textbf{H1b.}] Developers report higher perceived productivity gains when experiencing higher cognitive load while using AI assistants.\\
    \textit{Rationale: Higher cognitive engagement during AI interaction may lead developers to overestimate productivity improvements.}
    \item[\textbf{Experience and Cognitive Load}]
    \item[\textbf{H2a.}]Less experienced developers use AI coding assistants more frequently than more experienced developers for programming tasks.\\
    \textit{Rationale: Prior work suggests that less experienced developers may rely more heavily on AI assistance ~\cite{peng2023impact} Understanding this relationship helps identify which developer groups benefit most from AI tools in terms of productivity.}
    \item[\textbf{H2b.}]More experienced developers experience lower cognitive load when using AI tools compared to less experienced developers.\\
    \textit{Rationale:More experienced developers may have a stronger foundation for contextualizing AI assistance and have better mental models for evaluating AI output.}
    \item[\textbf{Familiarity and Cognitive Load}]
    \item[\textbf{H3.}]Developers experience lower cognitive load when using AI assistants to understand unfamiliar libraries compared to using traditional documentation.\\
    \textit{Rationale: AI assistants can provide contextual, natural language explanations and examples, potentially reducing the cognitive overhead of navigating and interpreting unfamiliar library documentation.}   
\end{description}

% \begin{description}
%     \item[\textbf{H1.}] There exists a negative correlation between perceived and actual productivity gains when using AI assistants for information seeking, while a positive correlation exists between cognitive load and perceived productivity gains.
    
%     \item[\textbf{H2.}] There exists a negative correlation between developer's years of experience and the frequency of AI tool use (or total time spent using AI tools). 
%     % \item[\textbf{H2.}] There exists correlations between developer expertise and use of AI assistants for coding tasks.
%     % \begin{description}
%     %     \item[\textbf{H2.1.}]  There is a positive correlation between a developer's prior knowledge in a domain and their rate of success when using AI assistants for information seeking in that domain.
%     %     \item[\textbf{H2.2.}]  There is a negative correlation between developer's years of experience and the frequency of AI tool use (or total time spent using AI tools)
%     % \end{description}
%     \item[\textbf{H3.}]  Developers experience a lower cognitive load using AI assistants for working with unfamiliar technologies.
%     \item[\textbf{H4.}] Developers need more validation effort when using AI assistants than when using traditional sources.
% \end{description}

\section{Variables}
% For each variable, you should give: – name (e.g., presence of family) – abbreviation (if you intend to use one) – description (whether the family of the crew members travels on board) – scale type (nominal: either the family is present or not) – operationalization (crew members without family on board vs. crew members with family onboard)
\subsection{Independent Variables}
Participants will be assigned to either use or not use AI-powered tools when seeking information for completing software development tasks. The \textbf{information seeking method (ISM)} will be a boolean variable \textit{(AI, no AI)}. \textbf{Developer knowledge (DK)} will be assessed through a pre-questionnaire where participants will report their experience with relevant libraries and programming concepts on a Likert scale \textit{(novice, intermediate, expert)}.
In terms of productivity, we will assess \textbf{actual productivity (AP)} using the following metrics: (1) time spent on information seeking (2) time spent writing code (3) time taken to complete the task. For assessing \textbf{perceived productivity (PP)}, participants will report these metrics in percentage in the post-questionnaire.
For H2, examining \textbf{expertise (Exp)} and \textbf{frequency of AI usage (t-AI)}, we will measure time spent using AI assistants as well as the number of prompts written, total number of tokens in the prompt(s), number of follow-up prompts, number of conversations.
For assessing \textbf{actual cognitive load (ACL)}, we will collect physiological measurements through eye tracking (pupil diameter) and EEG (alpha, beta, theta, gamma waves). For \textbf{perceived cognitive load (PCL)}, we will use scores reported on NASA-TLX.
% For measuring \textbf{validation effort (VE)}, we will track the (1) number of validation episodes, (2) resources consulted, and (3) time spent verifying solutions.

We will also control several \textit{confounding variables} in our study. The development environment will be standardized using a consistent VS Code setup with predefined extensions and dependencies. All participants will use JavaScript as the programming language and the session will be limited to one hour for the main task. The complexity and duration of the task will be guided by our pilot efforts to ensure it can be reasonably completed within the stipulated time and reduce fatigue effects. In addition, all participants will work with the same codebase to ensure code quality is controlled and does not impact productivity~\cite{cheng2022improves}, and task requirements and acceptance criteria will be standardized through unit tests. Environmental conditions will be controlled by conducting all sessions in the same quiet room with consistent temperature and lighting conditions. The hardware setup will be standardized across sessions, with all participants using the same laptop model, external monitor configuration, and input devices. EEG headset and eye tracker positioning will follow standardized placement protocols and be calibrated before each session. These controls will be validated during our pilot study to ensure they effectively minimize unwanted variation, with adjustments made based on pilot findings before proceeding with the main study.

% The total number of tasks for the study will be guided by our pilot efforts. If the number of main tasks is more than 1, we will randomize the task order across participants to control for learning effects.
% \eah{- Careful: Randomizing task order is not an adequate tool to mitigates learning effects. Learning effects stem from one subject participating in multiple tasks related tasks. Learning effects exits, no matter the order.}
% \subsubsection{Cognitive Load (CL)}
% \begin{itemize}
% \item Eye tracking metrics:
% \begin{itemize}
% \item Mean fixation duration
% \item Fixation count per minute
% \item Average saccade velocity
% \item Pupil size variation coefficient
% \end{itemize}
% \item EEG measurements:
% \begin{itemize}
% \item Alpha, beta, theta wave patterns
% \end{itemize}
% \item NASA-TLX scores
% \end{itemize}
% \subsubsection{Validation Effort (VE)}
% \begin{itemize}
% \item Count of validation episodes
% \item Number of resources consulted
% \item Time spent verifying (minutes)
% \end{itemize}
% \subsection{Control Variables}
% \begin{itemize}
% \item Development environment (standardized VS Code setup)
% \item Access to resources (controlled set of AI tools)
% \item Programming language (JavaScript)
% \item Session duration (maximum 1 hour per task)
% \item Task order
% \end{itemize}

\section{Materials and Tasks}
In this section, we outline the logistics of our study design which includes a pre- and post-questionnaires, study environment setup, and study session tasks. The study materials will be made available online once finalized.

\subsection{Pre-questionnaire}
We will require participants to complete a pre-questionnaire prior to participating in the study. 
The pre-questionnaire will collect participant demographic data including age, gender, education level, and current professional role to ensure a representative sample and enable analysis of potential demographic influences on AI-assisted tool usage. We will also collect data on participants' technical expertise (e.g., years of programming experience) and experience using AI-assisted tools for various software development tasks.

% \begin{itemize}
%     \item demographics
%     \item technical expertise
%     \item AI tool use experience
% \end{itemize}

% \subsection{Logistics}
\subsection{Experimental Setup}
To evaluate our hypotheses, we designed a controlled observation study which will involve hardware and software configurations that we discuss below.

\subsubsection{Data Collection Mechanisms}

For our study, we will collect both physiological and interaction data as participants complete each task.

\paragraph{Physiological Data} To collect physiological data, we will leverage hardware for measuring eye movements and brain activity during task completion. 
More specifically, following protocols and metrics outlined in prior work~\cite{hauser2018eye,peitek2022correlates}, 
%[CITE FRITZ]
we will use a Tobii Pro Fusion \footnote{\url{https://www.tobii.com/products/eye-trackers/screen-based/tobii-pro-fusion}} to collect: (1) gaze location, which tracks exactly where developers focus their attention; (2) fixation count and duration, which indicates specific elements developers focus on and how long; (3) saccade count and duration, which provide insights into how developers visually navigate between different elements to reveal context switches and comprehension patterns; and (4) pupil diameter, which is correlated with mental workload and processing demands.
    % We will use precise eye tracking measurements to capture developers' visual attention patterns and cognitive processes during software development tasks. Following established protocols from prior work [CITE FRITZ], we will use a Tobii Pro eye tracker to collect the following: (1) gaze location information to track exactly where developers focus their attention the process (2) fixation count and duration to measure where and how long developers concentrate on specific elements, indicating deeper cognitive processing (3) saccade count and duration to analyze how developers visually navigate between different elements, revealing their context switching and comprehension patterns, and (4) pupil diameter to assess cognitive load and mental effort as developers work through different programming tasks, since pupil dilation correlates with mental workload and processing demands. 

We will also capture brain activity for more in-depth cognitive load measurement. To measure brain activity, we will use the Emotiv EPOC X EEG headset\footnote{\url{https://www.emotiv.com/collections/all/products/epoc-x}} to generate an electroencephalogram for participants as they complete each task.
We selected this device for its higher accuracy with 14 channels and validated use in software engineering research for investigating attention in code review tasks~\cite{molleri2019experiences}.

We expect this approach to deliver more precise cognitive load measurements than single-channel alternatives like Neurosky MindWave~\footnote{\url{https://store.neurosky.com/pages/mindwave}}, especially for software development tasks that engage multiple cognitive processes simultaneously~\cite{fucci2019replication}.
Following best practices from prior EEG studies in software engineering, all sessions will be conducted in a quiet room with consistent lighting, controlled temperature with minimal electromagnetic interference and fixed monitor setup.

\paragraph{Interaction Data} To collect interaction data and help associate and contextualize the data collected via other mechanisms, we will use screen recording software to record the screen and audio during each session (i.e., Mac screen recorder). Prior work has leveraged screen and audio recordings to investigate various phenomena in software engineering-related research, including the effectiveness of tool recommendation styles~\cite{brown2020comparing}, questions asked while diagnosing security vulnerabilities~\cite{smith2015questions}, and small ``microtask'' contributions to larger source code repositories~\cite{latoza2018microtask}. This will allow us to track and conduct post-hoc analyses of participants' task completion efforts with and without AI tools.

% {Environment Recording}
% \begin{itemize}
%     \item Screen recording software
%     \item Keystroke logging
%     \item IDE interaction tracking
%     \item Resource access logging
% \end{itemize}

\subsubsection{Development Environment}

We will ask participants to complete one or more coding tasks (see Section~\ref{subsec:tasks}) in our experiment. To facilitate task completion, we will set up a laptop with an up-to-date installation of VS Code, which will include Git support out-of-the-box. This is important, given participants will be engaging with open source repositories during task completion. Participants will also be allowed to use the command line to access Git version control if preferred.
We will make sure that VS Code has all the necessary dependencies installed for each task.
All participants will also have access to a web browser, and those in the treatment condition will also have access to the following AI-assisted tools: GitHub Copilot,\footnote{\url{https://github.com/features/copilot}} ChatGPT,\footnote{\url{https://chatgpt.com/}} Claude,\footnote{\url{https://claude.ai/}} and Gemini.\footnote{\url{https://gemini.google.com/}} These large language model (LLM)-based systems are widely available either in VS Code or online and among the most popular generative AI developer assistants used in practice~\cite{haque2024information}. For the study, we will also connect the laptop to an external monitor for participants to complete study tasks with the eye tracker. We chose to have participants use a provided machine instead of their own laptop to ensure the environment is set up properly (i.e., dependencies) and control for a consistent environment across participants.
% \begin{itemize}
%     \item VS Code installed and set up with the necessary dependencies
%     \item Git version control \eah{Will this be inside of VS Code or via terminal?}
%     \item required dependencies
%     \item AI tools \textit{(GitHub Copilot, ChatGPT, Claude, Gemini)}
%     \item Web browser
% \end{itemize}
% \eah{Strengths from reviews: Usage of existing programmer tasks, striving for population diversity, detailed description of how measures are taken.  Pilot study intended. Sample size estimation intended, based on existing comparative findings. Great detailed description of measurements to be taken and adequate tool selection.}
\subsection{Training}
To ensure participants are proficient in using Visual Studio Code (VS Code) and to minimize the potential for unfamiliarity with the development environment to influence results, we will provide a quick walkthrough before the tasks begin. This session will introduce participants to key features of VS Code, including navigation through the file explorer, code editing, managing environments, and using the integrated terminal for executing commands. For participants in the experimental group, we will provide additional training on AI-assisted features, such as generating and refining code with GitHub Copilot. To allow participants to get comfortable with the interface, participants will complete a warm-up task as outlined below~\ref{subsec:tasks}. 
\subsection{Coding Tasks}\label{subsec:tasks}
Each session will engage participants in a series of technical tasks in our controlled environment consisting of a pre-configured development setup with all necessary dependencies and testing frameworks installed.

First, we will provide participants with a 15-minute \textbf{warm-up task} where they will solve a LeetCode-style algorithmic problem to get familiar with the development environment, available tools, and to ensure proper calibration of the eye-tracking and EEG equipment.

For the \textbf{main task} (approximately 1 hour), we will utilize open issues from popular open source repositories that are labeled as good for newcomers to their project, such as ``good first issue'' or ``beginner-friendly''. The selected issues must satisfy the following criteria:

\begin{enumerate}
    \item be reasonably completed within a one-hour timeframe to avoid fatigue effects
    \item require minimal domain-specific knowledge to isolate general problem-solving cognitive load from domain expertise effects
    \item contain a set of unit tests that serves as a validation criteria to enable measurement of solution correctness
\end{enumerate}

% % \subsubsection{Main Task (1 hour)}
% For the main portion of the observation study, we will utilize open issues from popular open-source repositories that are labeled as "good first issue" or "beginner-friendly". The selected issues must satisfy several criteria: (1) be reasonably completed within a one-hour timeframe to avoid fatigue effects, (2) require minimal domain-specific knowledge to isolate general problem-solving cognitive load from domain expertise effects, and (3) . 
We will source potential tasks from repositories listed in the ``awesome-for-beginners''\footnote{https://github.com/MunGell/awesome-for-beginners} GitHub collection, focusing on JavaScript projects, as it is the most widely used programming language according to Stack Overflow's 2024 Developer Survey.\footnote{\url{https://survey.stackoverflow.co/2024/technology/}}
To validate and finalize our experimental design, we will pilot the study with at least five developers to determine the optimal complexity and number of tasks, validate task completion time estimates, assess participant fatigue, and document common pitfalls and required clarifications.
% * task should be hour-long at most to avoid fatigue bias
\subsection{Labeling Session}
To complement the data collected through physiological and interaction mechanisms, we will conduct a labeling session where participants annotate their coding activities using the CodeRec User Programming States (CUPS) taxonomy~\cite{mozannar2024reading}. This session will enable us to analyze programmers’ behavior during AI-assisted coding tasks systematically.

Immediately following task completion, participants will review screen recordings of their sessions and annotate each segment with one of the following predefined CUPS states:
\begin{itemize}
    \item \textbf{Thinking/Verifying Suggestion:} Evaluating AI-generated suggestions for correctness or relevance.
    \item \textbf{Prompt Crafting:} Writing prompts or comments to influence AI-generated code.
    \item \textbf{Deferring Thought for Later:} Accepting suggestions with the intent to verify them at a later stage.
    \item \textbf{Thinking About New Code to Write:} Conceptualizing new functionality.
    \item \textbf{Writing New Functionality:} Implementing new code elements.
    \item \textbf{Editing Last Suggestion:} Modifying the most recently accepted AI-generated code.
    \item \textbf{Editing Written Code:} Adjusting code that was not derived from AI suggestions.
    \item \textbf{Debugging/Testing Code:} Investigating and resolving issues or running unit tests.
    \item \textbf{Looking up Documentation:} Referring to external resources for understanding code functionality.
    \item \textbf{Waiting for Suggestion:} Idle time awaiting AI-generated suggestions.
    \item \textbf{Writing Documentation:} Adding comments or docstrings for code clarity.
    \item \textbf{Not Thinking:} Periods of inactivity or distraction.
\end{itemize}
We will refine this taxonomy based on findings from our pilot study.
\subsection{Post-Questionnaire}
Following task completion, participants will complete a post-study questionnaire to capture their experience and assess various aspects of perceived cognitive load and tool usage. The primary instrument will be the NASA Task Load Index (NASA-TLX), a widely validated workload assessment tool~\cite{hart2006nasa}. 
The NASA-TLX measures six dimensions of workload: mental demand, physical demand, temporal demand, performance, effort, and frustration. Each dimension is rated on a 7-point scale with gradations from very low to very high (perfect to failure for the performance dimension). 

To evaluate the perceived role of AI assistance in development, participants will assess their interaction with the provided AI tools. This section will capture the perceived frequency of AI tool usage, perceived utility of AI suggestions and perceived productivity boost. These measures will help understand how developers integrate AI tools into their workflow and the impact on their cognitive processes.
The questionnaire will also include a task assessment section where participants evaluate their performance and experience. Participants will assess their familiarity with similar problems prior to the study, and describe specific challenges encountered during implementation. This self-assessment helps contextualize performance metrics and provides insight into how prior knowledge influences problem-solving approaches.

Environmental factors will be addressed through questions about the experimental setup and equipment. Participants will provide feedback on the comfort and potential interference of the EEG device during development, the usability of the provided development environment, and any environmental factors that may have affected their performance. This information is crucial for understanding any external influences on cognitive load measurements and ensuring the validity of our physiological data.
The entire questionnaire will be administered electronically immediately following task completion while the experience is fresh in participants' minds, reducing recall bias.

\section{Subjects}
% \eah{From reviews: - Are participants compensated ? It is unlikely to obtain a diverse population without adequate gratification, especially industrial developers are unlikely to participate without compensation.}
We will recruit participants through university mailing lists and our professional networks.
The target population includes graduate students and professional developers with at least one year of software development experience. To participate, each individual must be at least 18 years of age, familiar with JavaScript, and have no prior experience with the specific task domain. We will make efforts to recruit a diverse sample of participants across programming experience levels, professional roles, and other diversity axes---such as race/ethnicity, gender, education, and age.
Each participant will be compensated with a \$40 Amazon gift card upon successful completion.
% \eah{From reviews:  address their familiarity with the development environment (e.g., VS Code). This could impact the study’s validity, as unfamiliarity with the IDE might increase cognitive load or reduce productivity, regardless of task difficulty or AI assistance. I recommend the authors include familiarity with VS Code as a recruitment criterion or control variable.}
% \eah{From reviews: - If participants are filtered based on their IDE proficiency with only one IDE, this will most likely introduce a bias. Preferably would be to provide the required training and obtain the most diverse population group.}
% \textit{Inclusion Criteria:}
% \begin{itemize}
% \item at least 18 years of age
% \item software development experience
% \item familiarity with JavaScript programming
% \item no prior experience with the specific task domain
% \end{itemize}

We will conduct power analysis to determine the appropriate sample size needed for statistical significance and generalizability of results. This analysis will consider effect sizes from similar studies in software engineering research, desired statistical power (typically 0.8), and significance level (alpha = 0.05). The results of this analysis will guide our recruitment and ensure our findings have sufficient statistical validity.

% * need to do power analysis to determine sample size for generalizability
%https://www.spotfire.com/glossary/what-is-power-analysis
\section{Execution Plan}
We will obtain approval from our institutional review board (IRB) for human subjects research before commencing this experiment. Our study will employ a between-subjects design conducted in two phases: a pilot study followed by the main study. 
The pilot study will begin with the recruitment of at least five participants who match our target population criteria. 
These participants will undergo the complete experimental protocol, allowing us to validate our procedures, test measurement instruments, and gather initial feedback. 
Through this pilot phase, we will refine task instructions, adjust time allocations, determine number and complexity of tasks, improve our measurement instruments, and resolve any technical issues that arise. 

Following the pilot, we will begin participant recruitment for the main study. Each candidate will complete a pre-study questionnaire to assess their programming experience, demographic information, and typical development practices. To address concerns about the unreliability of self-reported data, we will include a short, standardized skill test as part of the recruitment process. This test will evaluate participants' knowledge on concept relevant to the task.
% \eah{IDE training}
% \eah{From reviews: - Self reported data is most likely not objective and hard to draw conclusions from. I would suggest to run a preliminary skill test, rather than relying on “years of experience”. In my experience self reported participant experience is unreliable.
% - If participants are filtered based on their IDE proficiency with only one IDE, this will most likely introduce a bias. Preferably would be to provide the required training and obtain the most diverse population group.}
Based on pre-questionnaire responses, we will select participants meeting our inclusion criteria. Participants will be randomly assigned to either the experimental or control condition using stratified randomization to ensure balanced distribution of experience levels between groups. The study will involve two groups:
\begin{itemize}
    \item \textbf{Experimental Group}: Participants will complete tasks using AI-enabled tools.
    \item \textbf{Control Group:} Participants will complete tasks without the aid of AI-enabled tools.
\end{itemize}
Participants in both groups will perform the same set of tasks to ensure consistency and comparability. Importantly, participants will remain within their assigned condition (AI-enabled or non-AI) for the entirety of their participation.

% \eah{from reviews: - I miss a detailed description of the study layout indicating how participant tasks are distributed. Implicitly it seems to me that there are two groups and participants either only deal with tasks without AI, or with tasks involving AI. From the description it is not clear if participants are assigned AI vs nonAI tools for the integrity of all their tasks, or per task.}
Each participant will complete the study protocol in the following sequence:
    \begin{enumerate}
        \item a 2-minute relaxation period using standardized mindfulness exercises to establish baseline cognitive state~\cite{fritz2016leveraging}
        \item a 15-minute warm-up programming task (LeetCode-style algorithmic problem) that allows participants to familiarize themselves with the development environment and tools
        \item a 1-hour main experimental task investigating and resolving the assigned software issue. The control group will work without AI assistance while the experimental group will have access to AI assistants. Throughout the session we will collect screen recordings, activity log, EEG and eye tracker readings, which will be used in data labeling and analysis
        \item labeling session to annotate the recording of their coding session using the CUPS taxonomy~\cite{mozannar2024reading}.
        \item an exit post-questionnaire which will allow us to gather data on perceived productivity and cognitive load, to be taken immediately after task completion (or the end of allocated time).
    \end{enumerate}
% \eah{From reviews: 10. The control group (working without AI coding assistants) seems primarily relevant for H4. However, it is unclear what "traditional sources" are provided to the control group, and how this setup will generate comparable data for other hypotheses. The authors should also elaborate on how findings from the control and experimental groups will be integrated across all hypotheses.}
The collected data will then be processed to extract our dependent variables, which will form the basis for our analysis.

% \begin{itemize}
% \item Recruit 5 participants for pilot
% \item Test experimental setup and procedures
% \item Refine tasks and measurement instruments based on feedback
% \end{itemize}

% \begin{itemize}
% \item Participant screening and scheduling
% \item Pre-study questionnaire administration
% \item Task completion sessions:
% \begin{itemize}
% \item Equipment setup and calibration
% \item Task briefing
% \item Task completion
% \item Post-task questionnaires
% \end{itemize}
% \item Post-study questionairre
% \end{itemize}

\section{Analysis Plan}
% \eah{from reviews: - From the description it seems that participants either use AI or no AI technology for their participation. I could not find a explicit description for how data will be paired.}
% \eah{From reviews: - The analysis plan in Section VIII generally makes sense, but lacks details on when a hypothesis will be rejected/corroborated. E.g., for H1, when is a correlation "large enough"? For an RR, I would expect these details. Similarly, alpha levels/p values should probably be mentioned, or at least how they will be determined.}
We will employ non-parametric statistical methods for our analysis as they make no assumptions about underlying data distributions and are stable even for small sample sizes. 
\subsection{H1: Cognitive Load vs. Perceived and Actual Productivity}
For H1a and H1b, examining correlations between perceived productivity, actual productivity, and cognitive load, we will first analyze the treatment group. We will extract perceived productivity scores from the post-task questionnaires, where participants will rate their productivity on a 7-point Likert scale. For actual productivity metrics, we will calculate task completion times in minutes, as well as the time spent on writing code and information seeking. To test H1a, which examines potential differences between perceived and actual productivity gains, we will calculate descriptive statistics including minimum, maximum, mean, and median values for both measures. The primary analysis will use a Wilcoxon Rank Sum test to compare perceived versus actual productivity gains. H1b will follow a similar analysis plan.

For H2a, analyzing the relationship between experience levels and AI tool usage, we will first categorize participants into experience levels (novice, intermediate, expert) based on our pre-study questionnaire. We will then calculate AI tool usage metrics, including the total time spent using AI tools and the frequency of AI tool interactions (i.e., number of prompts written, number of tokens in each prompt, number of turns in each conversation, number of conversations). We will use Spearman's rank correlation to test the relationship between experience level and these usage metrics, reporting correlation coefficients with bootstrap confidence intervals. Results will be visualized using box plots grouped by experience levels. H2b will also follow a similar analysis plan.

To assess H3, comparing cognitive load between AI assistant and traditional documentation conditions, we will employ a Wilcoxon Rank Sum test.

\section{Threats to Validity}
% \eah{Suggested by reviewers: [2] Moreno-Lumbreras, David, Jesus M. González Barahona, and Gregorio Robles. "Diving into Software Evolution: Virtual Reality vs. On-Screen." (2024).
% }
\subsection{Internal}
\begin{itemize}
    \item \textbf{Instrumentation Effects}: The EEG headset and eye tracking equipment may affect participants' natural behavior and increase cognitive load. We address this by including a warm-up period for participants to become comfortable with the equipment and by asking about equipment interference in our post-study questionnaire.
    \item \textbf{Environmental Factors}: Variations in room lighting, temperature, or external noise could affect physiological measurements. We aim to control for this by conducting all sessions in the same controlled environment with consistent conditions.
    \item \textbf{Fatigue Effects}: Extended use of the EEG headset or eye tracker might cause physical discomfort and affect performance. We limit the main task to one hour to minimize fatigue.
\end{itemize}
\subsection{External}
\begin{itemize}
\item \textbf{Programming Language}: Our focus on JavaScript may limit generalizability to other programming contexts. We chose JavaScript as it is widely used and allows us to recruit from a broader participant pool.
\item \textbf{AI Tool Selection}: We focus on specific AI tools (GitHub Copilot, ChatGPT, Claude, Gemini) which may not represent all AI-assisted development tools. We selected these tools as they are among the most widely used in practice.
\item \textbf{Environment Setting}: The controlled environment may not reflect developers' natural work settings. We attempt to mitigate this by using familiar development tools (VS Code).
\item \textbf{Participant Pool}: Our recruitment from university networks and professional contacts may not represent the full diversity of software developers. We address this by aiming to explicitly recruit across different experience levels and demographic backgrounds.
\end{itemize}
\subsection{Construct}
\begin{itemize}
    \item \textbf{Productivity Metrics}: Our chosen productivity metrics (task completion time, code writing time, information seeking time) may not capture all aspects of development productivity. We supplement these with qualitative data from post-task questionairre to provide a more complete picture.
    \item \textbf{Tool Familiarity}: Differences in participants' prior experience with VS Code or the provided AI tools could confound our results. We partially mitigate this through our warm-up task and training session, but acknowledge this as a limitation.
    \item \textbf{Task Complexity}: Our assessment of task complexity may not align with all participants' perceptions. We will validate our complexity assessments through pilot testing of selected tasks.
\end{itemize}

\bibliographystyle{ieeetr}
\bibliography{ref}

\end{document}